\documentclass[aps,prl,reprint,superscriptaddress,longbibliography]{revtex4-2}
\usepackage[T1]{fontenc}
\usepackage[utf8]{inputenc}
\usepackage{lmodern}
\usepackage{amsmath,amssymb,mathtools}
\usepackage{bm}
\usepackage{graphicx}
\usepackage{xcolor}
\usepackage{siunitx}
\usepackage{physics}
\usepackage{microtype}
\usepackage{float}
\graphicspath{{figures/}}
\usepackage[]{hyperref}
\usepackage[nameinlink,capitalize]{cleveref}

\begin{document}

\title{Janus Percolation in Anisotropic Limited-Degree Networks}

\author{Jacopo A. Garofalo}
\email{jacopoalexander.garofalo@unicampania.it}
\affiliation{Department of Mathematics and Physics, University of Campania “Luigi Vanvitelli”, Viale Lincoln 5, Caserta, 81100, Italy}
\author{Nuno A.M. Araújo}
\affiliation{Centro de Física Teórica e Computacional, Faculdade de Ciências, Universidade de Lisboa, Campo Grande 016, Lisboa, 1749-016, Portugal}
\author{Lucilla de Arcangelis}
\affiliation{Department of Mathematics and Physics, University of Campania “Luigi Vanvitelli”, Viale Lincoln 5, Caserta, 81100, Italy}
\author{Alessandro Sarracino}
\affiliation{Department of Engineering, University of Campania ‘‘Luigi Vanvitelli’’, Via Roma 29, Aversa, 81031, Italy}
\author{Eugenio Lippiello}
\affiliation{Department of Mathematics and Physics, University of Campania “Luigi Vanvitelli”, Viale Lincoln 5, Caserta, 81100, Italy}
\date{\today}

\begin{abstract}
Many real-world infrastructures, from sensor and road networks to power grids, are spatially embedded and anisotropic, with constraints on the maximum number of links each node can establish. Such systems can be represented as anisotropic limited-degree networks, in which each node forms at most $q$ outgoing links preferentially oriented along a fixed direction. By increasing the node density $\sigma$ at fixed $q$, we uncover a reentrant percolation transition: a giant strongly connected component emerges, but unexpectedly disintegrates again at high densities. This counterintuitive behavior implies that adding nodes, normally expected to enhance robustness, can instead reduce mutual accessibility and weaken global connectivity. The critical behavior displays two coexisting ``faces'': random-percolation scaling along the preferred direction and directed-percolation scaling transversely, therefore we name this phenomenon \emph{Janus percolation}, in analogy with the dual-faced Roman god. These findings demonstrate that anisotropy and degree limitation can jointly induce a novel reentrant connectivity with mixed universality that bridges the universality classes of random and directed percolation, providing fresh insight into how structural constraints shape connectivity and resilience in spatial networks.
\end{abstract}
\newpage
\maketitle


\textit{Introduction}. Many real-world network infrastructures are spatially embedded and subject to both anisotropy and degree constraints imposed by physical or operational limitations. In sensor networks, for example, ranked routing algorithms introduce a preferred direction for message transmission, while constraints on energy consumption and signal interference limit the number of active connections each node can sustain. Similar limitations in directional organization and connectivity arise in other large-scale systems such as power grids, communication, transportation, and road networks~\cite{kim2024shortest,karrer2014percolation,franceschetti2007closing,kleinberg2000navigation,gotesdyner2022percolation,buldyrev2010catastrophic,neta2024self,haas2002gossip,cogoni2021stability,verbavatz2021one,kocc2013matcasc,sensornetwork_application1}. These two features, anisotropy and limited degree of the node, also appear in many other real-world systems. 
Examples include Protein–Protein Interaction Networks~\cite{GKN08}, where molecular binding is directionally constrained and highly valence-limited, as well as social networks~\cite{FeetAL22}, in which individuals maintain only a finite number of ties and interactions are often inherently non-reciprocal. Clarifying how these constraints govern the architecture and resilience of complex networks is therefore a central question that spans physics, engineering, biology, and the broader study of complex systems.
\begin{figure}[b]
  \centering
  \includegraphics[width=\columnwidth,keepaspectratio]{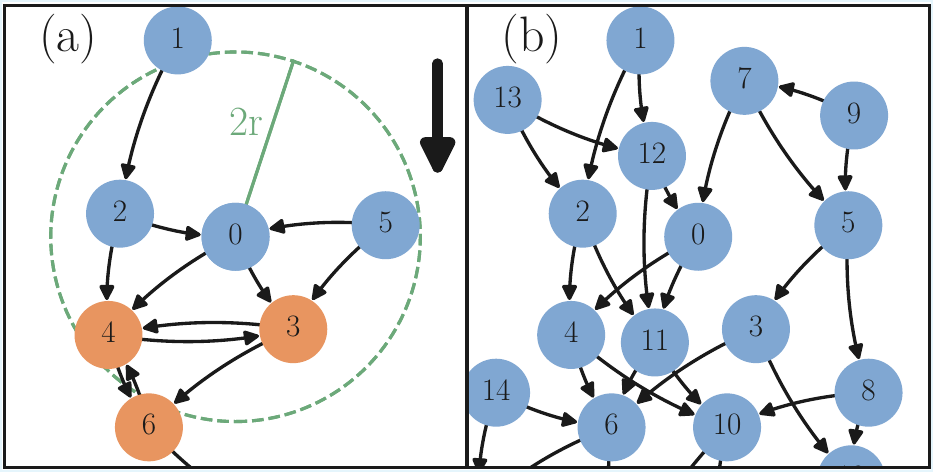}
  \caption{\textbf{From isotropic to anisotropic connectivity with increasing density.} Schematic illustration of the model for a maximum out-degree $q=2$, at density $\sigma=0.3$ (a) and $\sigma=0.6$ (b). Each node $i$ connects to up to $q$ neighboring nodes located within a distance $2r$, preferentially oriented downward along a fixed vertical direction (e.g node 0 connects to nodes 3 and 4, in panel a). We show a low density configuration in panel (a), and higher density of nodes in panel (b). Increasing density, more neighbors become available, allowing nodes to connect further along the preferred vertical direction (e.g. node 2 switching connection from node 0 in (a) to 11 in (b)), thus enhancing anisotropy. Connectivity is analyzed in terms of strongly and weakly connected components (SCCs and WCCs): in SCCs (orange) every node can reach every other by following directed links, whereas in WCCs (orange and blue) the same definitions holds when links are treated as undirected. Note that in panel (b), because of the strong anisotropy, SCC disappears.}
  \label{fig1}
\end{figure}
\begin{figure}[t]  
  \centering
\includegraphics[width=\columnwidth]{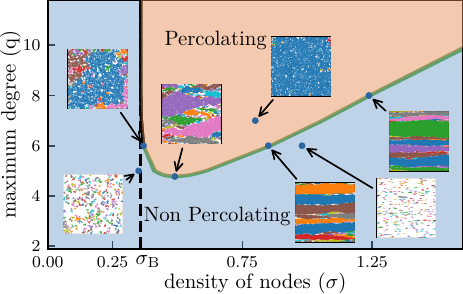}
  \caption{
\textbf{Reentrant percolation phase diagram of strongly connected components.} Phase diagram in the $(\sigma, q)$ plane, where $\sigma$ is the node density and $q$ the maximum out-degree. The solid line marks the transition between the non-percolating and percolating phases, where the black and green colors are used 
to denote  RP and JP universality classes, respectively. As $\sigma$ increases at fixed $q \gtrsim 4.7$, a giant strongly connected component (SCC) first emerges (percolating phase, orange) but this fragments again at higher densities (non-percolating phase, blue), revealing a reentrant transition. Insets show representative snapshots for a system of size $L=1600$, where colors denote distinct SCCs; only components with size $M>500$ are displayed. The vertical dashed line corresponds to the critical threshold ($\sigma_\mathrm{B}$) of the Boolean model.
  }
  \label{fig:phasediagram}
\end{figure}

Percolation, first introduced in Physics by Broadbent and Hammersley~\cite{broadbent1957percolation}, provides the prototypical framework to study how local connectivity rules give rise to large connected structures in disordered systems~\cite{araujo2014recent,saberi2015recent,dorogovtsev2008critical,cohen2000resilience}. At the percolation threshold, a macroscopic connected cluster emerges through a continuous phase transition characterized by universal critical exponents. For isotropic systems, these exponents fall into the universality class of random percolation (RP)~\cite{stauffer2018introduction,gawlinski1981continuum,meester1996continuum}, which describes connectivity transitions governed solely by geometry and randomness. When links or paths are strictly oriented along a specific direction, the system instead belongs to the directed percolation (DP) universality class~\cite{redner1981percolation,redner1982directed,janssen2000random,Marro_Dickman_1999,hinrichsen2000non,frey1994crossover}. DP exhibits self-affine scaling and, beyond geometric networks, captures the critical behavior of a wide range of nonequilibrium processes with anisotropic propagation. Together, RP and DP define the main universality classes for isotropic and anisotropic connectivity transitions in disordered media.

Here, we introduce a model of spatially embedded directed networks in which each node can establish at most $q$ outgoing links to neighbors within a given distance, preferentially oriented along a fixed direction. This rule mimics situations where connections preferentially follow a directional bias, such as gravity, flow, or information propagation, while remaining locally random. As illustrated in Fig.~\ref{fig1}, the degree of anisotropy depends sensitively on the density of nodes $\sigma$. At low densities (panel~(a)), few neighbors are available, therefore, nodes cannot consistently select the preferred orientation (top to bottom), leading to nearly isotropic connectivity. As density increases (panel~(b)), nodes have more potential neighbors to choose from, allowing links to align more effectively along the preferred direction. This interplay gives rise to a competition between connectivity, which favors more links, and anisotropy, which restricts their orientation. To quantify how this competition affects the global structure, we examine the formation of strongly and weakly connected components (SCCs and WCCs)~\cite{dorogovtsev2001giant,de2018percolation,schwartz2002percolation}. SCCs are defined as sets of nodes that can reach one another through directed links, representing regions of mutually accessible nodes, and thus robust bidirectional connectivity. WCCs, in contrast, are the sets of nodes that remain connected when link directions are ignored, capturing the broader backbone of the network. Analyzing the SCCs, we uncover a reentrant percolation diagram in the $(\sigma,q)$ phase space shown in Fig.~\ref{fig:phasediagram}. As density increases, a giant strongly connected cluster first emerges, but unexpectedly breaks apart again at higher densities. Finite-size scaling of the correlation lengths reveals a mixed critical behavior, with distinct exponents along the directions parallel and transverse to the preferred one: $\nu_{p} = 4/3$, consistent with RP, and $\nu_{t} =1.734$, characteristic of DP. We refer to this dual-scaling behavior as \emph{Janus percolation} (JP), in analogy with the two-faced Roman god, to emphasize the coexistence of two universality classes at the same transition.

\textit{Model}. The model is depicted in Fig.~\ref{fig1}. Following a classical Boolean model, we place $N$ nodes uniformly at random in a square domain of size $L\times L$ using a Poisson point process with open boundary conditions~\cite{meester1996continuum}. Two nodes $(i,j)$ are considered neighbors if their Euclidean distance satisfies $\|{\bf r}_j-{\bf r}_i\|<2r$, with ${\bf r}_i$ denoting the position of node $i$ and $r$ the maximum interaction range. We express all lengths in units of $r$ and explore system sizes ranging from $L=800$ to $L=12800$. 
To introduce directionality, for each node $i$ we define the list of neighboring nodes $\mathcal{L}_i=\bigl\{\, j\neq i \;\big|\; \|{\bf r}_j-{\bf r}_i\|<2r \,\bigr\}$. The elements of $\mathcal{L}_i$ are ordered by increasing $y$-coordinate (bottom to top), imposing a preferred vertical direction. Node $i$ then establishes directed outgoing links to the first $q$ entries of this ordered list, thus favoring neighbors with the lowest $y$-coordinate within the interaction range. The in-degree remains unconstrained. The way we defined our model, it is equivalent, for $q\to \infty$ to a Boolean model with a fixed radius $r$. For non-integer $q$, we define $q'$ and $p$ as its integer and decimal components (i.e. $q=q'+p$). Each node will have either $q$ or $q+1$ outgoing links with probability $1-p$ and $p$ respectively.
 SCCs are identified using Kosaraju’s algorithm~\cite{k-spda-10}, while WCCs are determined with a Hoshen–Kopelman procedure~\cite{hoshen1976percolation} applied to the same network after replacing all directed links by undirected ones. 
\begin{figure*}[t]
  \centering
  \includegraphics[width =\textwidth,keepaspectratio]{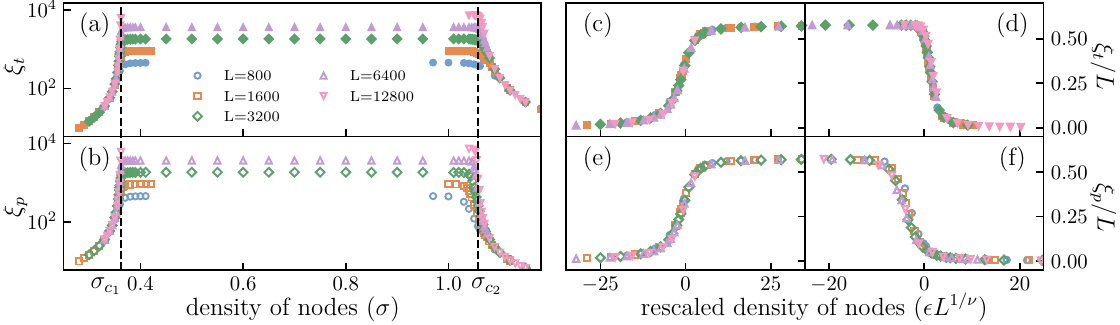}
  \caption{\textbf{Finite-size scaling of correlation lengths revealing a JP transition.} Filled and open symbols represent correlation lengths measured along the direction transverse (top panels) and parallel (bottom panels) to the preferred direction, respectively. Panels (a) and (b) show the raw correlation lengths for strongly connected components (SCCs) at $q=7$, the vertical dashed lines are the thresholds for the first and second transition. The corresponding data collapses are presented in the right column: (c,d) are respectively the collapses for the first and second transition of panel (a), similarly (e,f) are those for data in panel (b). Collapses are obtained using the finite-size scaling \textit{Ansatz} of Eq.~\ref{scaling_ansatz}, with $\epsilon = (\sigma - \sigma_{c_i})/\sigma_{c_i}$ with $i=1$ for panels (c,e) and $i=2$ for panels (d,f). The best data collapses yield $\sigma_{c_1} = 0.3615$ and exponents $\nu_{p} = \nu_{t}= 4/3$  for the first transition (panels~(c,e)), and $\sigma_{c_2} = 1.058$ with $\nu_{p} = 1.734$ and $\nu_{t}=4/3$ for the second transition (panels ~(d,f)).}
  \label{fig:correlations}
\end{figure*}
Since the preferred downward direction induces strong anisotropy, we measure correlation functions along the parallel and transverse axes. Let $g_{t}(x)$ denote the site-site correlation function, defined as the probability that two sites at a distance $x$ along the horizontal axis belong to the same connected component in the ensemble under consideration. We only consider sites that are within a vertical distance $2r$. An equivalent definition holds for $g_{p}(x)$. Averages are made over multiple independent realizations of the system. We then compute the correlation length as 
\begin{equation}
  \xi_{t}^2 \;=\; 
  \int_{0}^{L} x^{2} g_{t}(x)\ dx \Bigg/
       \int_{0}^{L} g_{t}(x) dx,
  \label{eq:xi_def}
\end{equation}
which characterizes the typical extent of connected regions along the direction transverse to the preferred one. An analogous definition applies to define  the correlation length $\xi_{p}$ that measures the size of the connected regions along the direction parallel to the preferred one. Close to the critical point, both $\xi_{t}$ and $\xi_{p}$ grow as a power law of the distance to the threshold, $\xi_z \sim |\sigma-\sigma_c|^{-\nu_z}$, where $z$ indicates ${t}$ or ${p}$, and $\sigma_c$ the critical threshold. In anisotropic systems it is possible that $\nu_p \ne \nu_t$.
In finite systems, this divergence is limited by the system size $L$, which introduces a natural cutoff and causes $\xi_z$ to depend on both $\sigma$ and $L$. This dependence is captured by the standard finite-size scaling {\it Ansatz} \cite{stauffer2018introduction},
\begin{equation}
\xi_{t}(\sigma,L) \sim L\,
\mathcal{F}_{t}\!\left[ \frac{(\sigma-\sigma_c)}{\sigma_c}\, L^{1/\nu_{t}} \right],
\label{scaling_ansatz}
\end{equation}
where $\mathcal{F}_t$ is the universal scaling function. A similar finite-scaling \textit{Ansatz} can be formulated for $\xi_{p}$.

\textit{Results}. To explore how anisotropy and limited degree affect global connectivity, we focus on the SCCs. Results for WCCs are provided in the Supplementary Material~\cite{SM}. We observe that, starting from small $\sigma$ and increasing its value, both $\xi_{t}$ and $\xi_{p}$ grow and reach a plateau at the threshold $\sigma_{c_1}$ (Fig.~\ref{fig:correlations}). 
Interestingly, $\xi_{t}$ remains approximately constant for $\sigma>\sigma_{c_1}$ until $\sigma$ exceeds a second threshold $\sigma_{c_2}$, beyond which it begins to decrease and eventually approaches zero at large $\sigma$. An analogous behaviour is observed in the parallel direction.
The plateau values $\hat{\xi}_{t}(L)$ and $\hat{\xi}_{p}(L)$ clearly depend on the system size $L$, and both increase approximately linearly with $L$. This behaviour is the hallmark of the onset of a percolation transition, where $\xi_{t}$ and $\xi_{p}$ diverge in the thermodynamic limit.
We perform a finite-size scaling analysis of the correlation lengths along both axes. 
The corresponding results are shown fixing $q=7$ in panels c, d, e, and f of Fig.~\ref{fig:correlations}. For the first transition, the data collapse is obtained using $\nu_{t}=\nu_{p}=4/3$, consistent with the correlation-length exponent of the RP universality class. In contrast, in the second transition the critical exponents differ along the two axes: along the transverse one, the best collapse is obtained with $\nu_{t} = \nu_{\parallel} \simeq 1.734$, characteristic of DP, while along the parallel one, the data are consistent with RP scaling, i.e. $\nu_{p} = 4/3$, instead of the typical $\nu_{\perp}\simeq 1.097$, of the DP universality class. To describe this transition is therefore necessary to introduce a new universality class (JP), which presents exponents of RP in one direction and DP in the perpendicular one. To verify the robustness of these findings, we present in Fig. 4 of the Supplementary Material \cite{SM} a comparison of data collapses obtained using alternative exponent sets for $\nu_{t}$ and $\nu_{p}$ (either RP or DP values). \\
We perform the same analysis for other $q$ values, and from finite-size scaling we identify points on the $(\sigma,q)$ plane (Fig.~\ref{fig:phasediagram}) where $\xi_{t}$ and $\xi_{p}$ diverge in the thermodynamic limit and others when they remain finite. For $q<q_c \simeq 4.7$ we find that the system is always in the non-percolating (blue) phase. By increasing $q$ two critical values of the density are observed $\sigma_{c_1}< \sigma_{c_2}$, such as that for $\sigma \in [\sigma_{c_1},\sigma_{c_2}]$ the system is always in percolating (orange) phase. 
The value of $\sigma_{c_2}$ increases by increasing $q$, indicating that the percolating thresholds extends up to $\sigma \to \infty$ in the large $q$ limit.
On the other hand, the value of $\sigma_{c_1}$ decreases by increasing $q$ up to reaching, for $q \gtrsim 6$, $\sigma_B\simeq0.359$ which is the percolation threshold of the Boolean model. In this large $q$-value interval the transition occurring at $\sigma_{c_1}$ is always in the RP universality class. Conversely, all other transition points manifest the mixed nature of the JP universality class.   We refer to the Supplementary Material for a finite-size scaling of the correlation lengths for $q=5$, for which the JP transition can be appreciated for both the first and second transitions \cite{SM}.
A precise determination of $q_c$ is challenging; however, our estimate is consistent with the minimum mean degree required for percolation in the classical isotropic Boolean model. In that case, indeed, percolation occurs only when the average node degree exceeds  the critical value $4.512$ \cite{meester1996continuum}.\\

The {\it reentrant} transition arises purely from the interplay between the degree constraint and the directional bias in link formation.
At low densities, the system consists of small disconnected clusters. As $\sigma$ increases, nodes find enough neighbors within the interaction range $r$ to form reciprocal connections, leading to the emergence of a giant SCC, which characterizes the percolating phase. However, at higher densities, the growing orientational bias of the links suppresses back-propagating connections: nodes increasingly link downward, leaving few available upward connections to sustain mutual reachability. Consequently, the giant SCC fragments again, marking a second transition back to the non-percolating phase.  Since reciprocal reachability is more easily maintained along directions where link orientation is isotropic, SCCs tend to break up preferentially along the direction parallel to the bias, while remaining more coherent transversely. This directional fragility explains the elongated morphology of SCCs observed in Fig.~\ref{fig:phasediagram} and in the snapshots shown in the insets.  
\begin{figure}[t]
  \centering
  \includegraphics[width=\columnwidth]{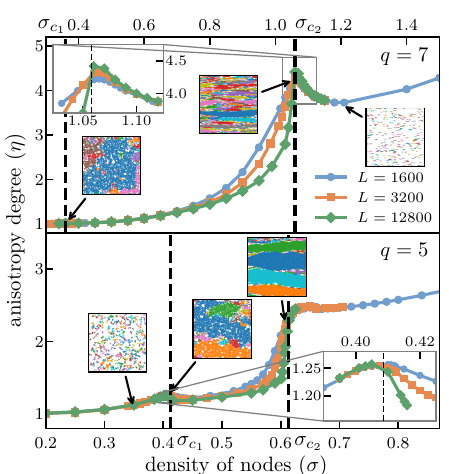}
  \caption{\textbf{Anisotropy degree of strongly connected components as a function of node density.} The anisotropy degree $\eta = G_{t}^2 / G_{p}^2$ quantifies the elongation of strongly connected components (SCCs), where $G_{t}$ and $G_{p}$ are the projected gyration radii along the transverse and parallel directions, defined in Eq.~\eqref{gyration}.The vertical dashed lines $\sigma_\mathrm{c_1}$ and $\sigma_\mathrm{c_2}$ are the thresholds for the first and second transitions. The top panel shows results for $q \ge 6$, where the first percolation transition is nearly isotropic, with $\eta \simeq 1$ (RP), while the second transition exhibits strong anisotropy with elongated SCCs (JP). The bottom panel corresponds to $4.7 \lesssim q < 6$, for which both transitions are JP. In all of these cases, $\eta$ reaches a pronounced maximum near the percolation threshold, signaling the formation of elongated clusters transverse to the preferred direction. }
  \label{fig: anisotropy}
\end{figure}

To quantify this anisotropic structure, we introduce an anisotropy degree, $\eta(\sigma,q)=\frac{\langle G_{t}^2\rangle}{\langle G_{p}^2\rangle}$, where $\langle \cdot\rangle$ denotes averages over realizations and $G_z^2$ is the square of the projected gyration radius along either transverse or parallel direction, defined as:
\begin{equation}\label{gyration}
G_{p} = \frac{1}{N_{\mathcal{C}}}\sum_{\alpha=1}^{N_{\mathcal{C}}} 
\left[ \frac{1}{|\mathcal{C}_{\alpha}|}\sum_{i\in \mathcal{C}_{\alpha}} (x_i - x_{\mathrm{av}})^2 \right]^{1/2} 
\end{equation}
Here, $N_{\mathcal{C}}$ is the total number of SCCs in a given realization, and $\mathcal{C}_{\alpha}$ the set of nodes belonging to the $\alpha$-th SCC. $G_t$ can be defined similarly. Fig.~\ref{fig: anisotropy} shows how $\eta$ varies with $\sigma$ and the two types of transition are clearly observed. In the case $q=7$, we find for the first transition $\eta \simeq 1$, that corresponds to the isotropic transition (black line of the phase diagram). The second transition, that is a JP transition, exhibits a pronounced peak in $\eta$ , reflecting the emergence of highly elongated components transverse to the preferred direction (green line in the phase diagram). In the case $q=5$ shown in Fig.~\ref{fig: anisotropy}, we have a further confirmation of the JP nature of both the first and second transitions for $q<6$, explained by the fact that $q$ is small enough so that even for small densities, at which the first percolating transition appears, there is enough anisotropy in the clusters to induce a JP transition.

\textit{Final remarks.} The results presented here reveal that combining geometric anisotropy with a limitation on the number of outgoing links profoundly alters the nature of connectivity in spatial networks. 
Even in the absence of disorder or temporal dynamics, these two simple ingredients are sufficient to produce a reentrant percolation transition: as node density increases, a giant strongly connected component first forms and then breaks up again at higher densities. 
This counterintuitive behavior originates from the competition between connectivity, which benefits from denser neighborhoods, and anisotropy, which constrains the available reciprocal links required for strong connectivity. 
Our findings thus demonstrate that increasing local connectivity does not necessarily enhance global robustness. This insight is directly relevant to engineered systems such as sensor networks, transport infrastructures, and communication grids, where directional constraints and limited capacity are ubiquitous.

Finite-size scaling further reveals that the reentrant transition cannot be described by a single universality class. 
Instead, it exhibits a mixed critical behavior, which we refer to as JP, with RP exponents along the preferred direction and DP exponents transversely. 
This mixed scaling suggests that anisotropy does not merely shift the critical threshold but fundamentally changes the structure of the transition, thereby bridging the RP and DP universality classes.
Future work should explore whether the JP transition observed here persists under alternative directional rules, or in the presence of disorder, and assess whether JP represents a broader class of mixed connectivity transitions in nonequilibrium or anisotropic media.

\begin{acknowledgments}
We akwnoledge the support from CINECA that provided computational resources through the ISCRA~C project \textit{IsCc7}.
We acknowledge financial support from the Portuguese Foundation for Science and Technology (FCT) under Contracts no. UID/00618/2025. EL \& LdA acknowledges funding from the MUR program PRIN 2022 PNRR P202247YKL.
\end{acknowledgments}
\bibliographystyle{apsrev4-2}
\bibliography{biblio}

\end{document}